\documentclass[journal]{IEEEtran}

\ifCLASSINFOpdf
\else
\fi

\usepackage{multirow}
\usepackage{graphicx}
\usepackage{makecell}

\begin{document}

%
\title{Data-Flow-Based Extension of the System-Theoretic \\
Process Analysis for Security (STPA-Sec)}
%

%
\author{}

\author{Jinghua~Yu,~Stefan~Wagner,~\IEEEmembership{Member~IEEE,}
        and~Feng~Luo
\thanks{J. Yu is with School of Automotive Studies, Tongji University, 201804 Shanghai China (e-mail: yujinghua@outlook.com).}
\thanks{S. Wagner is with Institute of Software Technology, University of Stuttgart, 70569 Stuttgart, Germany (e-mail: stefan.wagner@iste.uni-stuttgart.de)}
\thanks{F. Luo is with the School of Automotive Studies, Tongji University, 201804 Shanghai, China  (e-mail: luo\_feng@tongji.edu.cn)}
\thanks{This research was supported by the China Scholarship Council and funds of the German Federal Ministry of Education and Research under grant number 16KIS0995.}
\thanks{This version has been submitted to IEEE System Journal on 22 May 2020.}}

\maketitle

\begin{abstract}
Security analysis is an essential activity in security engineering to identify potential system vulnerabilities and achieve security requirements in the early design phases. Due to the increasing complexity of modern systems, traditional approaches, which only consider component failures and simple cause-and-effect linkages, lack the power to identify insecure incidents caused by complex interactions among physical systems, human and social entities. By contrast, a top-down System-Theoretic Process Analysis for Security (STPA-Sec) approach views losses as resulting from interactions, focuses on controlling system vulnerabilities instead of external threats and is applicable for complex socio-technical systems. In this paper, we proposed an extension of STPA-Sec based on data flow structures to overcome STPA-Sec’s limitations and achieve security constraints of information-critical systems systematically. We analyzed a Bluetooth digital key system of a vehicle by using both the proposed and the original approach to investigate the relationship and differences between both approaches as well as their applicability and highlights. To conclude, the proposed approach can identify more information-related problems with technical details and be used with other STPA-based approaches to co-design systems in multi-disciplines under the unified STPA process framework.
\end{abstract}


%

\section{Introduction}
%
%
%
%
\IEEEPARstart{S}{ystem} security is an emergent property of the system, which represents a state or condition that is free from asset loss and the resulting loss consequences. System security engineering, as a special discipline of system engineering, coordinates and directs various engineering specialties to provide a fully integrated, system-level perspective of system security and helps to ensure the application of appropriate security principles and methodologies during the system life cycle for asset protection \cite{ross2016nist}. Violating system security constraints causes unexpected incidents, like mission failure or leaking sensitive information, and finally leads to financial or even life losses. Therefore, security needs to be considered carefully in system design. Security analysis, referring to the activity of analyzing systems in security-related aspects to achieve security requirements in this research, is performed in the early security engineering phase and helps to manage system risks and make decisions.

Traditional security analysis approaches, being designed for former relatively simple systems, are not effective to analyze increasingly complex systems. Today's Cyber-Physical Systems (CPS) or Socio-Technical Systems (STS) consist of not only physical components but also software and even social elements, in which components in multi-disciplines interact with each other deeply. For example, an autonomous vehicle (as a CPS) consists of tens of thousands of physical components as well as lines of codes. A vehicle Over-The-Air (OTA) software update system (as an STS) refers to not only the technical part but also social entities like data providers and regulations. However, most traditional approaches start with system decomposition and focus on component failures, which leads to overlooking impacts of interactions since components are analyzed individually. Traditional causality models attribute accidents to an initial component failure cascading through a set of other components (like dominos) \cite{young2014inside} and can not address causes of losses with non-linear cause-and-effect linkages.

To meet the requirements of modern systems, a relatively new approach for safety engineering called System-Theoretic Process Analysis (STPA) was proposed \cite{leveson2019stpa} and the extension for security named STPA-Sec was presented later \cite{young2013systems}. However, we recognized some limitations of STPA-Sec when implementing it, especially for data-flow-based systems. Therefore, this research aims to work out an extended approach based on the unified STPA process framework for complex information-critical systems to overcome the identified limitations of STPA-Sec.

The rest of this paper is organized as follows. In section \uppercase\expandafter{\romannumeral2}, we introduce traditional approaches and the STPA series with research gaps and our contributions. In section \uppercase\expandafter{\romannumeral3}, we introduce the original STPA-based approaches and propose the extension in detail. In section \uppercase\expandafter{\romannumeral4}, we present the analysis process of a use case by using both original and extended approaches to demonstrate how to use them and make the comparison. Finally, we summary this paper in section \uppercase\expandafter{\romannumeral5}.

 




\section{Related Work}

\subsection{Traditional Security Analysis Approaches}

The purpose of the security analysis in the early design stage is to achieve security requirements. The Threats Analysis and Risk Assessment (TARA) is normally the main activity in the security analysis to identify potential threats and assess threat risks \cite{sae2016j3061}. The SAE (Society of Automotive Engineers) J3061 - Cybersecurity Guidebook for Cyber-physical Vehicle systems \cite{sae2016j3061} provides a list of approaches, which contain complete process frameworks of TARA, like the EVITA approach \cite{ruddle2009security} and TVRA \cite{etsi2017102}. Besides, techniques are proposed for only threat identification or risk assessment and can be used in the previously mentioned TARA frameworks, including Attack Tree Analysis (ATA) \cite{schneier2000secrets}, STRIDE threat models \cite{MS2009STRIDE}, Threat and Operability Analysis (THROP) \cite{sae2016j3061}, Threat Matrix \cite{mccarthy2014characterization} and BDMP-based (Boolean logic Driven Markov Processes) Modelling \cite{pietre2010modeling} for threat identification, as well as Binary Risk Analysis (BRA) \cite{BRA2011} and NIST SP 800-30 Guide \cite{ross2012guide} for risk assessment. Other than methods originally invented for security, some approaches are evolved from the safety field by introducing security awareness into the process and support the co-analysis of both safety and security. 

Table \ref{T_trad_approach} summarizes both original and evolved approaches with brief introductions. They are all bottom-up approaches building upon physical or functional decomposition instead of analyzing the system as a whole initially. They focus more on the tactic level and may overlook issues at the strategy level. The tactics are means to accomplish a specific action and focused on physical threats, while the strategy is regarded as the art of gaining continuing advantages and is focused on abstract outcomes \cite{young2014inside}. The latter one is useful to broaden the mind and includes more aspects like organizational and managerial ones in the analysis.

\begin{table*}[!t]
\renewcommand{\arraystretch}{1.1}
\caption{Summary of Traditional Security Analysis Approaches}
\label{T_trad_approach}
\centering
\begin{tabular}{p{3cm}<{\raggedright}p{5cm}<{\raggedright}p{9cm}<{\raggedright}}
\hline
\textbf{Type} & \textbf{Approach} & \textbf{Brief Introduction}\\
\hline
\multirow{4}{3cm}{Original Approaches with complete process frameworks} & E-Safety Vehicle Intrusion Protected Applications (EVITA) & EVITA approach considers four security objectives (safety, privacy, financial, operational) and uses attacks trees to identify threats and assess risks \cite{ruddle2009security}. \\
\cline{2-3} & Threat, Vulnerabilities, and implementation Risks Analysis (TVRA) & TVRA is a process-driven TARA approach to systematically identify unwanted incidents which need to be avoided \cite{etsi2017102}. \\
\cline{2-3} & Operationally Critical Threat, Asset, and Vulnerability Evaluation (OCTAVE) & OCTAVE is a process-driven TARA method which is best suited for enterprise information security risk assessments \cite{alberts1999operationally}. \\
\cline{2-3} & HEAling Vulnerabilities to ENhance Software Security and Safety (HEAVENS) & HEAVENS is a systematic approach of deriving security requirements for vehicle E/E systems, including processes and tools supporting for TARA \cite{HEAVENS2016}. \\
\hline
\multirow{3}{3cm}{Approaches evolved from other disciplines and support co-analysis} & A Security-Aware Hazard and Risk Analysis Method (SAHARA) & SAHARA is a combined approach of the Hazard Analysis and Risk Assessment (HARA) with the STRIDE model and outlines the impacts of security issues on safety concepts \cite{macher2015sahara}. \\
\cline{2-3} & Failure Mode, Vulnerabilities and Effects Analysis (FMVEA) & FMVEA is an approach evolved from the Failure Mode and Effect Analysis (FMEA) to identify vulnerability cause-effect chains for security \cite{schmittner2014security}. \\
\cline{2-3} & Combined Harm Assessment of Safety and Security (CHASSIS) & CHASSIS is a unified process for safety and security by using UML-based models (e.g. misuse cases and sequence diagrams) \cite{raspotnig2012combined}. \\
\hline
\end{tabular}
\end{table*}

\subsection{System-Theoretic Process Analysis (STPA) Approach and Extensions}

To overcome the limitations of traditional approaches, STPA was created as a hazard analysis approach based on the System-Theoretic Accident Model and Process (STAMP), which views losses as results from interactions among various system roles that lead to violations of safety constraints and analyzes issues at the strategy level. STPA provides a powerful way to deal with complexity by using traceable hierarchical abstraction and refinement \cite{young2014inside}.

Other than safety engineering, STPA has also been extended into other fields with the same system-theoretic thought. Young and Leveson \cite{young2013systems} firstly presented STPA for Security (STPA-Sec), which shares similar steps with STPA and focuses on controlling system vulnerabilities instead of avoiding threats. To perform co-analysis of safety and security under the STPA framework better, Friedberg et al. \cite{friedberg2017stpa} proposed a novel analysis methodology called STPA-SafeSec, which integrated STPA and STPA-Sec into one concise framework and overcomes limitations of original approaches (e.g. no considerations about non-safety security issues) by introducing security constraints and mapping abstract control structures to real components. Shapiro \cite{shapiro2016privacy} proposed STPA for Privacy (STPA-Priv), which extends STPA into privacy engineering by introducing privacy concepts and consideration into the general STPA process steps.

The most significant highlight of STPA-based approaches is that they shift from focusing on preventing failures and avoiding threats to enforcing safety constraints and control system vulnerabilities. Identifying and controlling system vulnerabilities rather than brainstorming and reacting to threats is a more efficient way to achieve system safety and security, because controlling a vulnerability may be effective to reduce several threats. Besides, threats are dynamic. Newly emerged threats can not always be detected in time, but controlling vulnerabilities can protect the system against even unknown threats, just like defending a castle by reinforcing walls and not caring who is the enemy. Another highlight is that the STPA-based approaches are applicable for socio-technical systems, which are systems that consider requirements spanning hardware, software, personal, and community aspects \cite{STSonline}. The analysis scope of the STPA series includes not only physical system components but also humans, natural or social environment and their interactions, which makes the approaches more suitable for today’s complex systems. Furthermore, due to the numbers of extensions of STPA in various disciplines, it is easier to perform co-design with similar STPA framework and the same system model.

\subsection{STPA-Sec Applications and Gaps}

The STPA-Sec approach or its extensions have been used to identify system security constraints in various industries. Khan, Madnick and Moulton \cite{khan2018cybersafety} demonstrated the implementation of STPA-Sec to identify security vulnerabilities of a use case (Central Utilities Plant Gas Turbine) in industrial control systems. Carter \cite{carter2018systems} used STPA-Sec with a previous information elicitation process to analyze a small reconnaissance unmanned aerial vehicles. Sidhu \cite{sidhu2018application} applied an STPA-Sec extension with modified attack tree method to analyze cybersecurity of autonomous mining systems. Wei and Madnick \cite{wei2018system} analyzed a use case (Over-The-Air software update) in the automotive industry by using both STPA-Sec and CHASSIS and compared analysis outcomes, which showed that STPA-Sec can identify more hazards compared to CHASSIS, while CHASSIS is more suitable for information lifecycle analysis. 

Nevertheless, researchers also pointed out several limitations of STPA-Sec. Schmittner, Ma and Puschner \cite{schmittner2016limitation} reported that the original STPA-Sec lacks guidance for intended causal scenarios, excludes considerations of the data exchange which is not directly connected to the process control and cannot cover more information-security centric properties such as confidentiality. Torkildson, Li and Johnsen \cite{torkildson2019improving} also found that some essential security issues can be overlooked and recommended to strengthen STPA-Sec by combining it with data-flow-based threat models. However, Torkildson's approach converts the STPA control structure into a data flow diagram by simply replacing control actions and feedback paths with data channels. Although such a data flow diagram helps to identify more data-related threats than using STPA-Sec alone, this diagram based on the original control loop does not view the system from the data aspect initially and may also miss data-related information. Besides, the STPA-Sec approach regards the security issue as one of the key threats affecting system safety \cite{wei2018system} and only supports the identification of safety-related security goals \cite{martin2017safety}. Non-safety-related security issues like confidentiality may be overlooked.

To sum up, existing STPA-Sec is not effective to identify non-safety but information-related issues since it does not consider security from the perspective of data flows. Furthermore, STPA-Sec lacks guidance for identifying security concepts.

\subsection{Contributions}
In this paper, we propose a data-flow-based extension of STPA-Sec (named STPA-DFSec) with elicitation guide words to overcome STPA-Sec's limitations. The analysis process of a vehicle digital key system is presented to demonstrate how to use STPA-DFSec. We also analyze the same system by using the original STPA-Sec and compare outcomes though synthesis and mapping. Finally, we discover the relationship between both approaches and conclude the highlights and applicability of them.

\section{Methodology}

\subsection{Brief Introduction of STPA and STPA-Sec}

We briefly introduce the original STPA framework as the basis of the proposed approach in this section. 

STPA starts with defining the purpose of the analysis, including system-level losses, hazards and constraints. Losses are about something valuable and unacceptable to the stakeholders. A hazard is a system state or set of conditions that, together with a particular set of worst-case environmental conditions, will lead to a loss. Finally, system constraints can be derived from identified hazards, which specifies system conditions or behaviors that need to be satisfied to prevent hazards and ultimately prevent losses \cite{leveson2019stpa}.

\begin{figure}[!ht]
\centering
\includegraphics[width=1in]{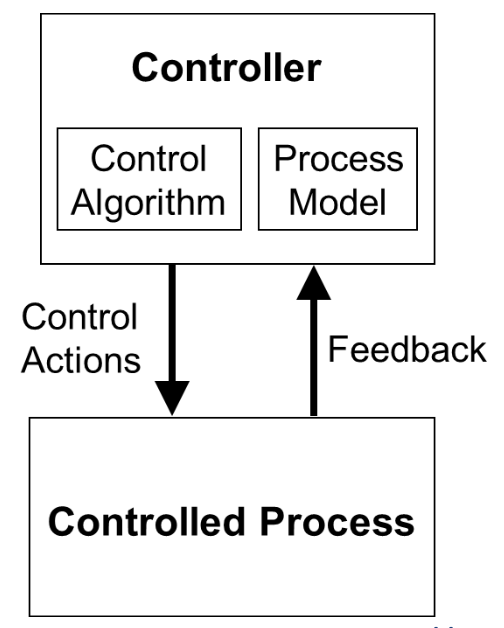}
\caption{General Control Loop \cite{leveson2019stpa}}
\label{F_ctrl_loop}
\end{figure}

Then, the control structure needs to be built to describe relationships and interactions by modelling the system as a set of control loops (show in Figure \ref{F_ctrl_loop}).

The third step is to identify unsafe control actions, which will lead to a hazard in a particular context and worst-case environment \cite{leveson2019stpa}, based on the previously built structure. Four ways of being unsafe are provided in STPA as guide words for the identification (listed in Table \ref{T_step_summary}).

Finally, loss scenarios, which describe the causal factors that can lead to unsafe control actions, will be identified. Two types of loss scenarios must be considered, which are ‘scenarios that lead to unsafe control actions’ and ‘scenarios in which control actions are improperly executed or not executed’ \cite{leveson2019stpa}. Each identified scenario represents a system failure which needs to be controlled by designers.

STPA-Sec, as the extension for security considerations, shares the same basic steps. Vulnerabilities, instead of hazards are identified in the first step since vulnerabilities lead to security incidents, which is just like hazards lead to safety incidents \cite{young2013systems}. Similarly, final identified loss scenarios represent system vulnerabilities which need to be controlled.

\subsection{STPA-DFSec Approach}

The proposed STPA-DFSec follows the general STPA framework but introduces a data-flow-based structure for information security considerations. The main steps are described as follows.

\subsubsection{Step 1 - Define the purpose of the analysis}

Just being similar with the STPA-Sec, the first step of the analysis is to identify system-level losses, vulnerabilities and constraints to figure out what are unacceptable results that need to be avoided at the system strategy level.

General security attributes like Confidentiality, Integrity and Availability (C.I.A. Triad Model) are guide words for the vulnerability identification, which classify the security problems and elicit potential vulnerabilities. 

Furthermore, to help to identify losses, STPA-DFSec provides general guidance for loss identification based on the study of various safety- and security-related definitions from standards and technical documents in industries including ISO 26262 \cite{ISO262622018}, EVITA project report \cite{ruddle2009security} and J3061 guideline \cite{sae2016j3061}. All possibilities of losses at a high abstract level are listed in Table \ref{T_gen_loss}. Losses of a particular case are a subset of this general list and can be described concretely according to practical situations.
{}
\begin{table}[!ht]
\renewcommand{\arraystretch}{1.3}
\caption{General List of Losses}
\label{T_gen_loss}
\centering
\begin{tabular}{p{0.5cm}p{2.4cm}<{\raggedright}p{4.3cm}}
\hline
\textbf{Label} & \textbf{Loss} & \textbf{Description}\\
\hline
L-1 & Loss of life or cause injury to life & Includes human and animal life \\
\hline
L-2 & Loss of physical property & Represents physical objects belonging to stakeholders (e.g. devices) \\
\hline
L-3 & Loss of non-physical property & Represents virtual properties belonging to stakeholders (e.g. sensitive information, reputation) \\
\hline 
L-4 & Loss of environment & Includes the natural or artificial world \\
\hline 
\end{tabular}
\end{table}

\subsubsection{Step 2 - Model Functional Interaction Structure}

\begin{figure}[!ht]
\centering
\includegraphics[width=2.5in]{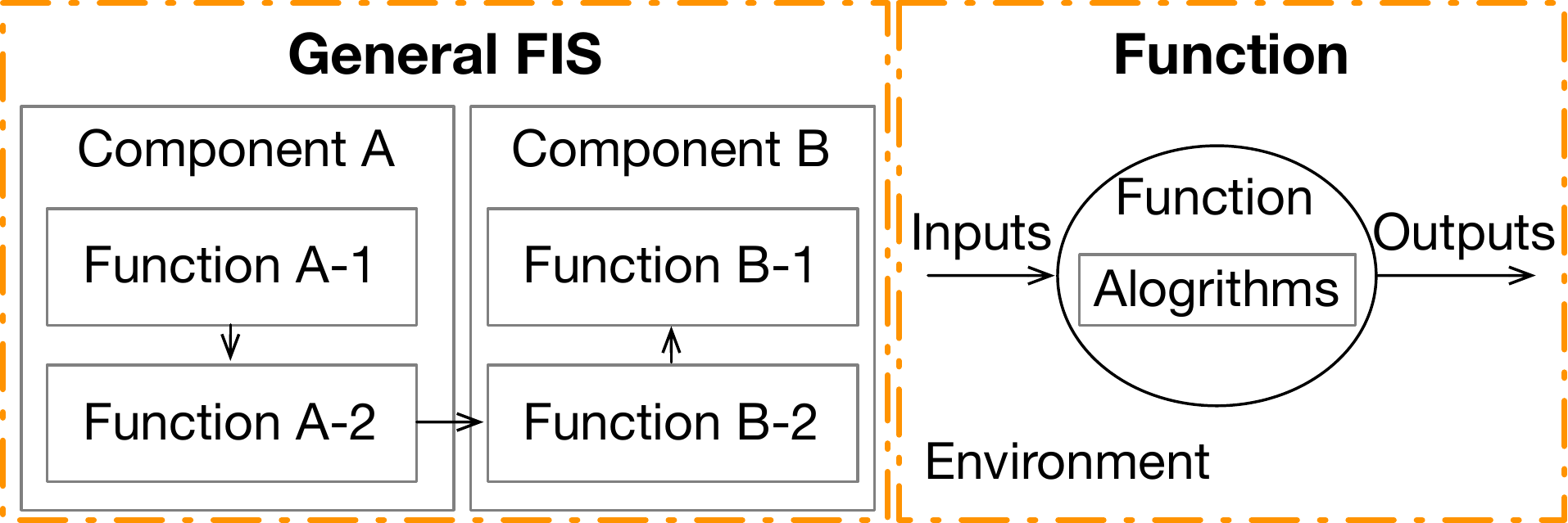}
\caption{General FIS and Element 'Function' }
\label{F_gen_FIS}
\end{figure}

Instead of the control structure, a Functional Interaction Structure (FIS) based on data flows is created to interpret how the system works from the perspective of data flows. The basic element of the FIS is the ‘Function’, which works based on its inputs and algorithms inside and outputs process results. The processing environment, together with inputs and algorithms, will affect function behaviors and final outputs. Inputs and outputs, instead of control actions and feedback, are interactions between components in FIS. Figure \ref{F_gen_FIS} shows a general interaction structure and the function element, in which arrows represents data flows.

\subsubsection{Step 3 - Identify Insecure Function Behaviors}

Based on the FIS, we can identify Insecure Function Behaviors (IFB) for the target system by following the basic technique in STPA. Insecure behaviors are identified with the help of Guide Words (GW), which are slightly modified to fit the proposed approach. Table \ref{T_IFB_temp} is the template for identifying IFBs.

\begin{table}[!ht]
\renewcommand{\arraystretch}{1.5}
\caption{Template for Identifying IFBs}
\label{T_IFB_temp}
\centering
\begin{tabular}{p{0.8cm}<{\raggedright}p{2.1cm}<{\raggedright}p{2.1cm}<{\raggedright}p{1.4cm}<{\raggedright}}
\hline
\textbf{Function (F)} & \textbf{GW:} Not being Executed Causes Vulnerabilities (NECV)$^1$ & \textbf{GW:} being Executed Causes Vulnerabilities (ECV)$^2$ & \textbf{GW:} Timing Issues (TI)$^3$ \\
\hline
S\_F$_n$ & S\_F$_n$\_IFB$_{m1}$$^4$ & S\_F$_n$\_IFB$_{m2}$ & S\_F$_n$\_IFB$_{m3}$ \\
\hline
\multicolumn{4}{l}{
\makecell[l]{\textbf{Notes:}\\
$^1$ Represents that the function is not executed successfully. \\
$^2$ Represents that the function is executed with improper conditions \\ (e.g. incorrect inputs or algorithms, process with data leakage risks). \\
$^3$ Represents that the execution exceeds the timing limits.\\
$^4$ S\_F$_n$\_IFB$_{m}$ is the label of each IFB, in which S represents the \\ subject of the function.}} \\
\hline
\end{tabular}
\end{table}

\subsubsection{Step 4 - Identify Loss Scenarios}

Finally, Loss Scenarios (LS), as possible causes of IFBs, are identified by using optimized guide words. Table \ref{T_LS_temp} is the template for identifying LSs with two classes of guide words. The ‘Function itself’ class helps to identify scenarios caused by unexpected behaviors inside the function, while the ‘Execution environment (Env)’ class refers to external conditions outside. 

\begin{table}[!ht]
\renewcommand{\arraystretch}{1.3}
\caption{Template for Identifying LSs}
\label{T_LS_temp}
\centering
\begin{tabular}{p{0.5cm}p{1.1cm}<{\raggedright}p{1.2cm}<{\raggedright}p{1.2cm}<{\raggedright}p{1.3cm}<{\raggedright}p{1cm}<{\raggedright}}
\hline
\textbf{IFBs} & \textbf{GW:} Function Itself & \textbf{GW:} Env- Function Inputs & \textbf{GW:} Env- Calling Behaviors & \textbf{GW:} Env- Computing Resources & \textbf{GW:} Env- Links\\
\hline
S\_F$_n$\_ IFB$_m$ & 
S\_F$_n$\_ IFB$_m$\_LS$_{p1}$ & 
S\_F$_n$\_ IFB$_m$\_LS$_{p2}$ & 
S\_F$_n$\_ IFB$_m$\_LS$_{p3}$ & 
S\_F$_n$\_ IFB$_m$\_LS$_{p4}$ & 
S\_F$_n$\_ IFB$_m$\_LS$_{p5}$ \\
\hline 
\end{tabular}
\end{table}

Each loss scenario represents a system vulnerability which should be controlled by designers or operators. Detailed system constraints can also be derived from loss scenarios by simply inversing the conditions of loss scenarios or defining what the system must do in case the incident occurs \cite{leveson2019stpa}. System constraints are inputs of further design phases.

\subsection{Summary}

Table \ref{T_step_summary} summarizes the process steps of both STPA-DFSec and STPA-Sec with highlights of differences, in which '+' donates added features of the STPA-DFSec and '*’ represents modified steps in comparison with the original STPA-Sec.

\begin{table*}[!t]
\renewcommand{\arraystretch}{1.3}
\caption{Summary of STPA-DFSec and STPA-Sec Steps}
\label{T_step_summary}
\centering
\begin{tabular}{p{2.8cm}p{5.4cm}p{5.2cm}}
\hline
\textbf{Basic Four Steps} & \textbf{STPA-DFSec Details} & \textbf{STPA-Sec Details} \\
\hline
Step 1 - Define the purpose of the analysis & 
Identify system-level losses, vulnerabilities and constraints, link vulnerabilities with corresponding losses and security attributes$^{+}$. A general losses list is provided$^{+}$.
& 
Identify system-level losses, vulnerabilities and constraints.\\
\hline 
Step 2 - Model the system structure & 
Model the system by functional interaction structure based on data flows$^{*}$. & Model the system by functional control structure based on control loops. \\
\hline 
Step 3 - Identify insecure items & Use modified guide words$^{*}$ (‘not being executed’, ‘being executed’ and ‘timing issues’) to identify insecure function behaviors. & Use guide words (‘not providing’, ‘providing’, ‘too early, too late, out of order’, ‘stopped too soon, applied too long’) to identify insecure control actions. \\
\hline 
Step 4 - Identify loss scenarios & Use modified guide words$^{*}$ (‘function itself’, ‘execution environment(incl. function inputs, calling behaviors, computing resources and links)’ to identify loss scenarios.
& Use guide words (‘unsafe controller behavior’, ’inadequate feedback and information’, ‘involving the control path’, ‘related to the controlled process’) to identify loss scenarios. \\
\hline 

\end{tabular}
\end{table*}

\section{Case Study}

\subsection{Use Case Definition and Assumption}

\begin{figure*}[!ht]
\centering
\includegraphics[width=6in]{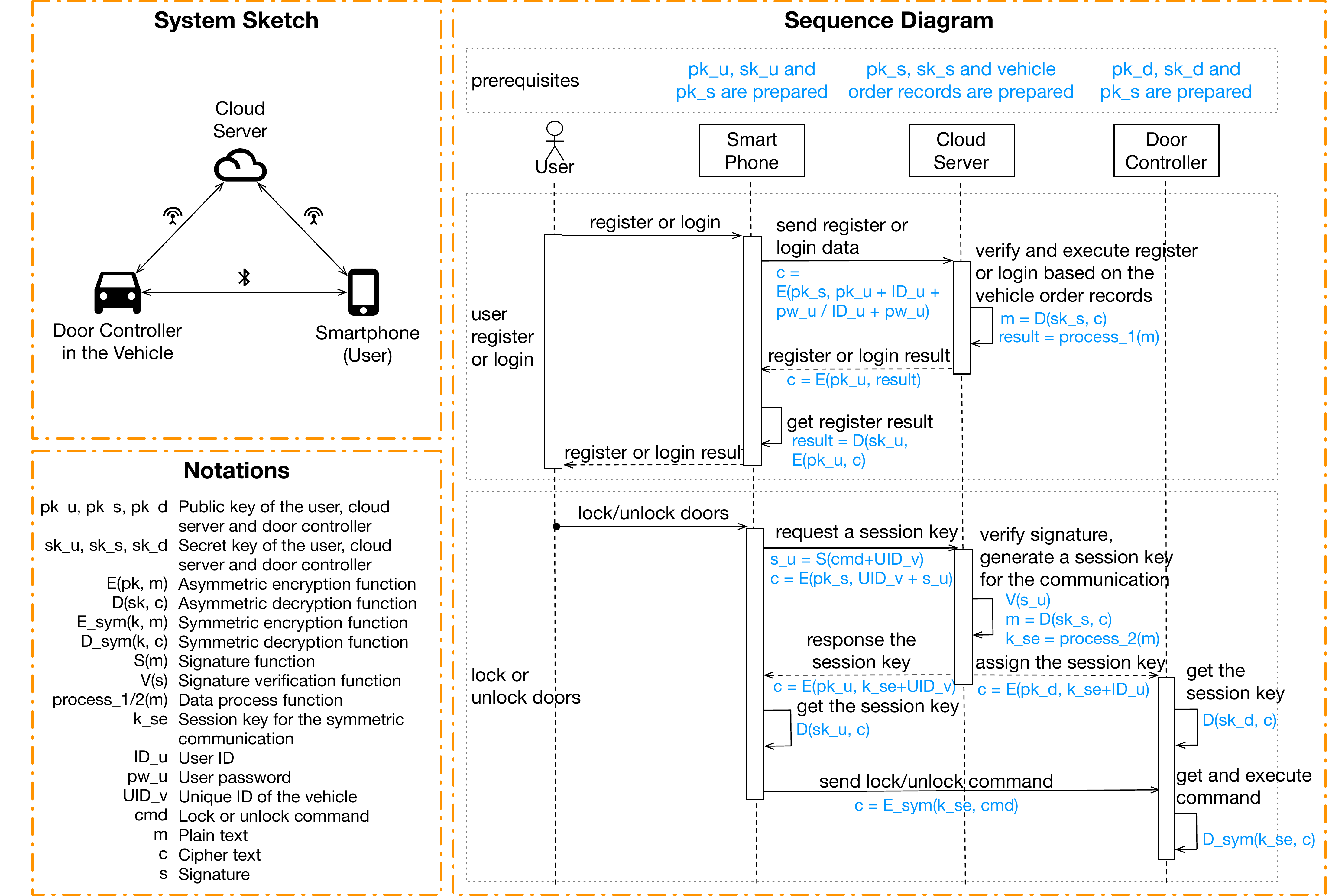}
\caption{Sequence Diagram of the System}
\label{F_SD}
\end{figure*}

In this section, a Bluetooth digital key system of a vehicle is introduced as the target system in this research. The system consists of three main physical components and aims to lock or unlock vehicle doors by using smartphones. Communication between different entities are through wireless channels and protected by cryptographic mechanisms. The system sketch and sequence diagram of two main services are shown in Figure \ref{F_SD} to describe how this system works.

In this example case, we assume that the connections between components have been established via Wi-Fi or Bluetooth in advance, but the connection is not ensured to be secure, and the prerequisites in Figure \ref{F_SD} are regarded trusted. In this research, we only focus on security issues, which means that the system can work properly without intended external disturbances and the system development errors and hardware random failures are out of scope. 

\subsection{Analysis by STPA-DFSec}

The analysis of the target system by using STPA-DFSec is presented in this section. First, system-level Losses (L), Vulnerabilities (V) with linked losses and security attributes and derived System Constraints (SC) are listed in Table \ref{T_ana_step1}. 

\begin{table}[!ht]
\renewcommand{\arraystretch}{1.3}
\caption{Losses, Vulnerabilities and Constraints of the Use Case}
\label{T_ana_step1}
\centering
\begin{tabular}{p{8cm}}
\hline
L-1: Loss of physical property (The vehicle and properties in it) \\
L-2: Loss of non-physical property (Manufacturer’s reputation) \\
\hline
V-1: Fail to lock the vehicle without being detected. [L-1/2, Integrity] \\
V-2: Fail to lock or unlock the vehicle, which can be detected by the user. [L-2, Availability] \\
V-3: Leak sensitive information. [L-1/2, Confidentiality] \\
\hline 
SC-1: Missions should be completed successfully. [V-1/2]\\
SC-2: If the mission fails, it must be detected by the user. [V-1]\\
SC-3: Sensitive information should be protected from leaking. [V-3]\\
SC-4: If sensitive information is leaked, it should be detected and reactions need to be taken to minimize losses. [V-3] \\

\hline 
\end{tabular}
\end{table}

\begin{figure}[!ht]
\centering
\includegraphics[width=3.3in]{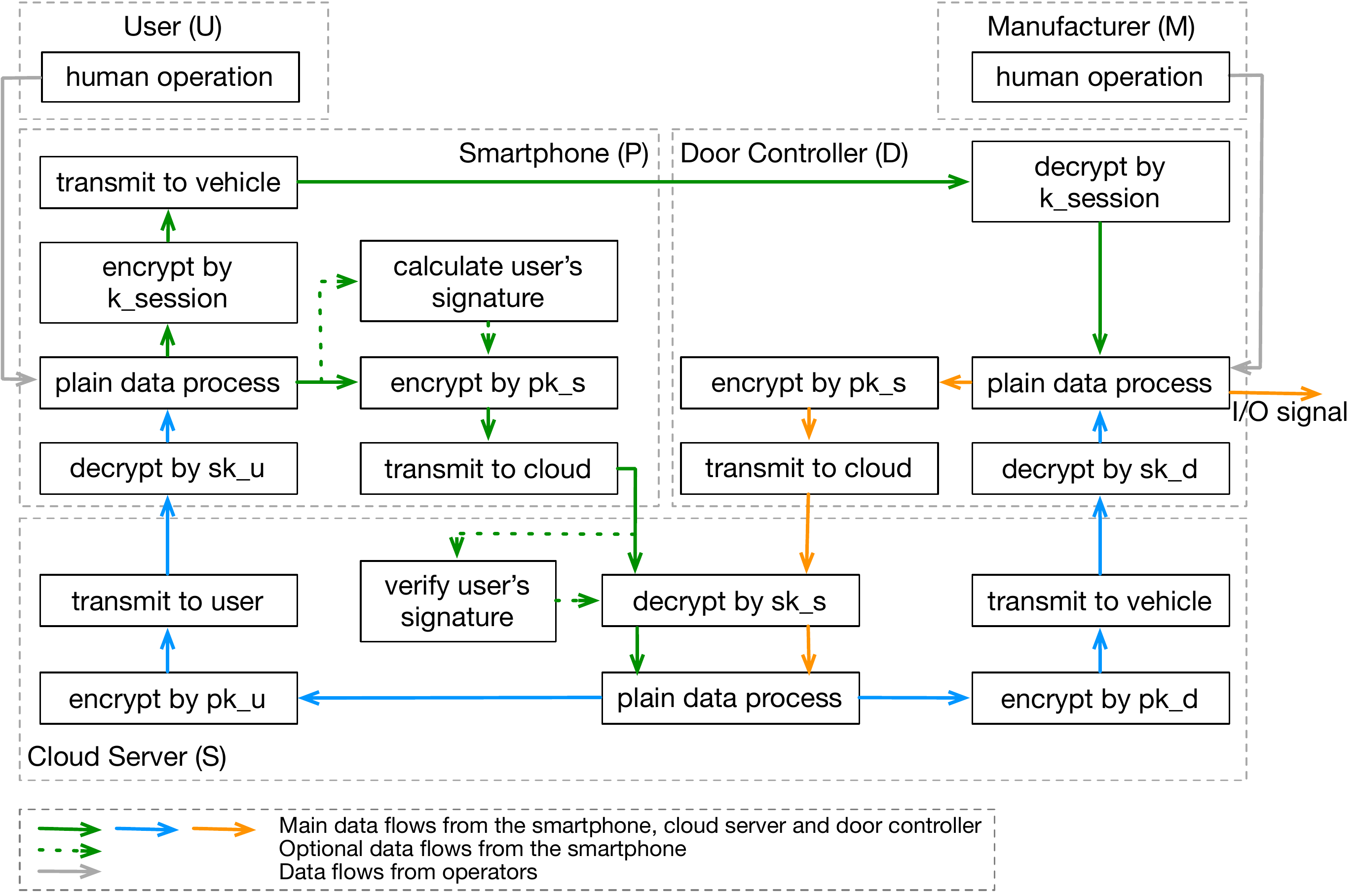}
\caption{Functional Interaction Structure based on Data Flows}
\label{F_FIS}
\end{figure}

Second, the functional interaction structure is created in Figure \ref{F_FIS} based on the system data flows. Two functions with identified IFBs are shown in Table \ref{T_IFB_part} as examples. In contrast to most traditional approaches, this analysis includes participants (user and manufacturer) outside the physical system boundary. Functions in a human operator can also be refined into more detailed movements like 'decision in the mind', 'pressing button' or 'recording password'. Since we focus on the physical part in this analysis, human movements are simplified as one ‘human operation’ function. Note that the first letter of the IFB labels in Table \ref{T_IFB_part} represents system components including smartphone (P), cloud server (S), door controller (D), user (U) and manufacturer (M).

\begin{table}[!ht]
\renewcommand{\arraystretch}{1.5}
\caption{Identified Insecure Function Behaviors}
\label{T_IFB_part}
\centering
\begin{tabular}{p{2cm}<{\raggedright}p{1.5cm}<{\raggedright}p{1.7cm}<{\raggedright}p{1.2cm}<{\raggedright}}
\hline
\textbf{F} & \textbf{GW: }NECV & \textbf{GW: }ECV & \textbf{GW: }TI \\
\hline
P/S/D\_F2: Data transmission & P/S/D\_F2 \_IFB1 & P/S/D\_F2\_ IFB2 & P/S/D\_F2\_ IFB3 \\
\hline
U/M\_F7: human operation & / & U/M\_F7\_IFB1, U/M\_F7\_IFB2 & /\\
\hline
\multicolumn{4}{l}{\textbf{IFB Description:}}\\
\multicolumn{4}{l}{
\makecell[l]{
P/S/D\_F2\_IFB1: The required data fails to be transmitted. [V-1/2] \\
P/S/D\_F2\_IFB2: The data is attacked (e.g. eavesdrop, tamper)\\ during the transmission. [V-1/2/3] \\
P/S/D\_F2\_IFB3: Function execution violates the timing limits. [V-2] \\
U/M\_F7\_IFB1: The invalid entity operates with valid data. [V-1/2] \\ 
U/M\_F7\_IFB2: The valid entity operates with invalid or \\ incorrect data. [V-1/2]
}}\\
\hline
\end{tabular}
\end{table}

Finally, LSs are identified for each IFB. Example LSs with related guide words in the bracket are listed in Table \ref{T_IFB_LS_part}.

\begin{table}[!ht]
\renewcommand{\arraystretch}{1.5}
\caption{Loss Scenarios of IFBs}
\label{T_IFB_LS_part}
\centering
\begin{tabular}{p{1.8cm}<{\raggedright}p{2.3cm}<{\raggedright}p{2.3cm}<{\raggedright}}
\hline
\textbf{IFBs} & \textbf{GW: }Function Itself & \textbf{GW: }Environment\\
\hline
P/S/D\_F2\_IFB3 & P/S/D\_F2\_IFB3\_LS1 & P/S/D\_F2\_IFB3\_LS2, P/S/D\_F2\_IFB3\_LS3, P/S/D\_F2\_IFB3\_LS4 \\
\hline
U/M\_F7\_IFB2 &U/M\_F7\_IFB2\_LS1 & / \\
\hline
\multicolumn{3}{l}{\textbf{LS Description:}}\\
\multicolumn{3}{l}{
\makecell[l]{
P/S/D\_F2\_IFB3\_LS1: The function algorithm is modified to \\cause corresponding insecure function behaviors. \\
P/S/D\_F2\_IFB3\_LS2: The input data size is modified, which \\requires more time to compute. (Env -Function Inputs) \\
P/S/D\_F2\_IFB3\_LS3: Computing resource is occupied to cause \\violation of execution timing limitations. (Env -Computing Resources) \\
P/S/D\_F2\_IFB3\_LS4: Transmission is slowed down by additional \\mechanisms on links (e.g. additional switches). (Env -Links) \\
U/M\_F7\_IFB2\_LS1: The valid entity is fooled intendedly to use \\invalid or incorrect data. 
}}\\
\hline
\end{tabular}
\end{table}

Note that the IFBs and LSs of various subjects are merged to make the list concise since different system components may contain the same functions and different insecure behaviors may have the same causalities. In practice, it is also meaningful to describe each function or LS separately to achieve security constraints for corresponding engineers, who might only have access to a part of design information due to security reasons.

\subsection{Analysis by STPA-Sec}

\begin{figure}[!ht]
\centering
\includegraphics[width=3.3in]{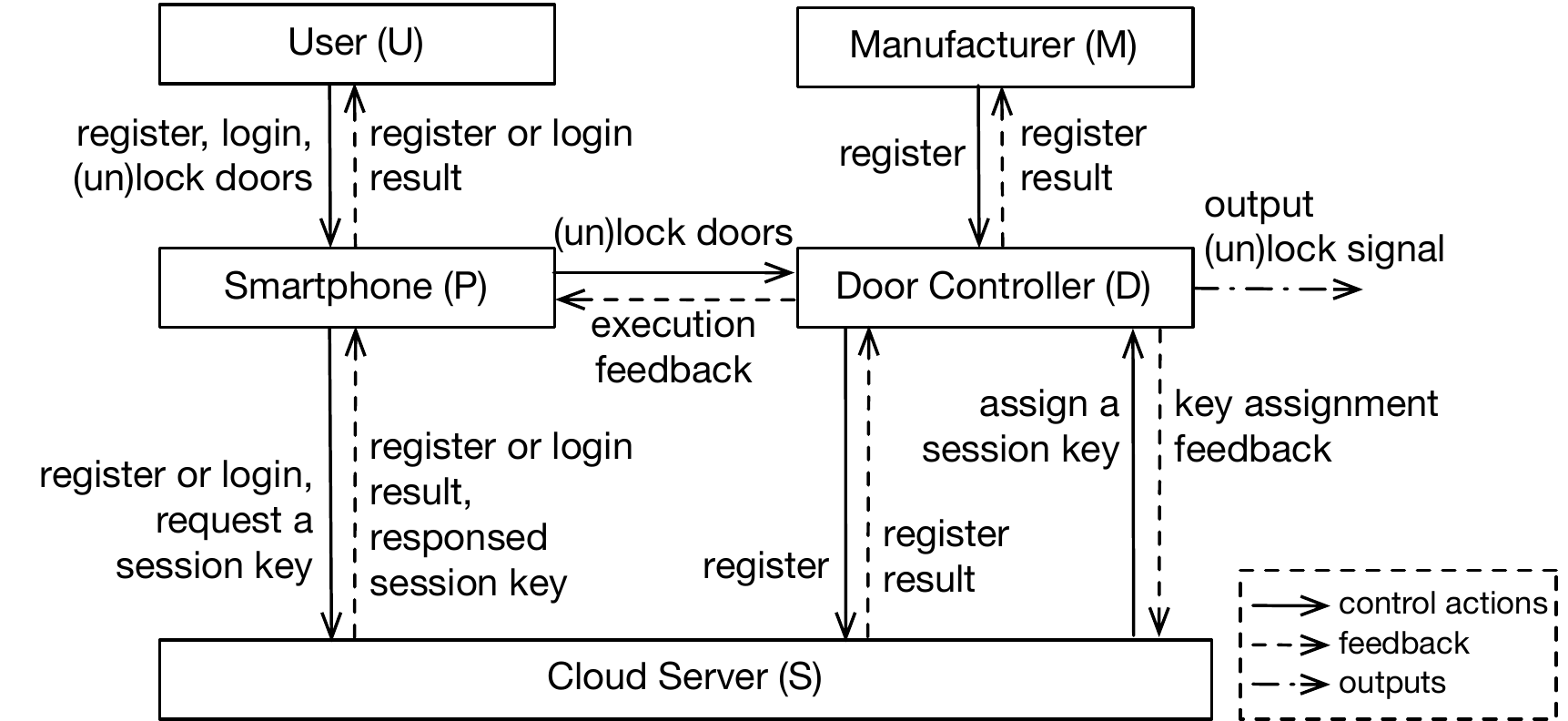}
\caption{Control Structure of the System}
\label{F_CAS}
\end{figure}

We also analyzed this use case by STPA-Sec. Due to the same system model, the identified losses, hazards and system constraints are the same as those in the STPA-DFSec analysis. Therefore, the work here starts with drawing the system control structure shown in Figure \ref{F_CAS}, and then Insecure Control Actions (ICA) are identified. ICAs of example Control Actions (CA) are shown Table \ref{T_ICA_part}, in which the first letter of the label represents the control action's controller.

\begin{table}[!ht]
\renewcommand{\arraystretch}{1.5}
\caption{Identified Insecure Control Actions}
\label{T_ICA_part}
\centering
\begin{tabular}{p{2cm}<{\raggedright}p{1.5cm}<{\raggedright}p{1.5cm}<{\raggedright}p{1.5cm}<{\raggedright}}
\hline
\textbf{CA} & \textbf{GW: }Not Providing & \textbf{GW: }Providing & \textbf{GW: }Timing Issues$^1$ \\
\hline
D\_CA1: Register & D\_CA1\_ICA1& D\_CA1\_ICA2& D\_CA1\_ICA3\\
\hline
U\_CA2: Lock or unlock doors & / & U\_CA2\_ICA1 & / \\
\hline 
\multicolumn{4}{l}{\textbf{ICA Description:}}\\
\multicolumn{4}{l}{
\makecell[l]{
D\_CA1\_ICA1: The door controller fails to register at the server. [V-2] \\
D\_CA1\_ICA2: The door controller registers with data leakage risks. \\ (e.g. UID, public key leakage). [V-3] \\
D\_CA1\_ICA3: The action violates the timing limits. [V-1] \\
U\_CA2\_ICA1: The user is fooled to send wrong commands. [V-1/2] \\
}}\\
\hline 
\multicolumn{4}{l}{
\makecell[l]{
\textbf{Notes:}\\
$^1$: Two GWs about timing in STPA-Sec are merged as ‘timing issues’.
}}\\
\hline

\end{tabular}
\end{table}

Finally, loss scenarios of each ICAs are identified. Some example LSs are shown in Table \ref{T_ICA_LS_part}.

\begin{table}[!ht]
\renewcommand{\arraystretch}{1.5}
\caption{Loss Scenarios of ICAs}
\label{T_ICA_LS_part}
\centering
\begin{tabular}{p{0.8cm}<{\raggedright}p{1.3cm}<{\raggedright}p{1.3cm}<{\raggedright}p{1.3cm}<{\raggedright}p{1.3cm}<{\raggedright}}
\hline
\textbf{ICAs} & \textbf{GW: }Controller& \textbf{GW: }Controller Path & \textbf{GW: }Controlled Process & \textbf{GW: }Feedback  Path \\
\hline
D\_CA1\_ ICA3 & D\_CA1\_ ICA3\_LS1 & D\_CA1\_ ICA3\_LS2 & D\_CA1\_ ICA3\_LS3 \\
\hline
U\_CA2\_ ICA1 & / & / & U\_CA2\_ ICA1\_LS1 & U\_CA2\_ ICA1\_LS2 \\
\hline
\multicolumn{5}{l}{\textbf{LS Description:}}\\
\multicolumn{5}{l}{
\makecell[l]{
D\_CA1\_ICA3\_LS1: The algorithm at the controller is tampered, \\which requires more calculating time.\\
D\_CA1\_ICA3\_LS2: The link is modified to slow down the \\ data transmission.\\
D\_CA1\_ICA3\_LS3: The algorithm at the controlled process is \\ tampered, which requires more calculating time.\\
U\_CA2\_ICA1\_LS1: The user interface of the smartphone is \\modified to guide the user to send wrong commands.\\
U\_CA2\_ICA1\_LS2: The vehicle state is presented wrongly, \\which leads to the user’s wrong behaviors. \\
}}\\
\hline
\end{tabular}
\end{table}

\subsection{Outcome Comparison}

The functions and control actions are basic elements in the STPA-DFSec and STPA-Sec respectively. Normally, a control action includes several functions to provide a service. For example, the control action D\_CA1 (Door controller registers at the server) consists of functions of data process, transmision, encryption and decryption. Therefore, the relationship between these two elements is that a sequence of the execution of functions forms a control action. 

To find out how different approaches work on the same use case, we mapped the analysis outcomes in both analyses. For example, D\_CA1\_ICA3\_LS1 and D\_CA1\_ICA3\_LS3 in the STPA-Sec analysis can be mapped to P/S/D\_F2\_IFB3\_LS1. U\_CA2\_ICA1\_LS1 and U\_CA2\_ICA1\_LS2 can be mapped to U\_F7\_IFB2\_LS1. After performing the mapping for all analysis outcomes, we found that each loss scenario in the STPA-Sec analysis can be mapped to a corresponding one in the STPA-DFSec analysis from the perspective of data process, which means that the proposed approach can find all possibilities identified by the original approach. Furthermore, more loss scenarios can be identified in the STPA-DFSec analysis. For example, the scenario 'Computing resource is occupied to cause violation of execution timing limitations.' (P/S/D\_F2\_IFB3\_LS3 in Table \ref{T_IFB_LS_part}) cannot be identified by the STPA-Sec. Therefore, more technical details related to the data process can be revealed by the data-flow-based analysis.

Another finding is that an STPA-DFSec loss scenario can be mapped to several STPA-Sec ones because a function is always called by various control actions for different applications. This explains why STPA-Sec can reveal more detailed information from the perspective of applications.

\subsection{Discussion}

After the comparison, we concluded the differences and highlights of both approaches. The STPA-DFSec focuses on information flows and discusses possible vulnerabilities along the whole data flow paths, which helps to identify more detailed loss scenarios from the perspective of information flows. By contrast, since control actions in STPA-Sec are derived from system functionalities, STPA-Sec can reveal more insecure details linked to concrete application scenarios. STPA-DFSec addresses where (in which function) a loss scenario occurs, while STPA-Sec addresses when (in which application scenario) a loss scenario occurs. 

Since both approaches have different advantages, how to choose an approach depends on particular cases. Two principles can be used to help the decision. The first one is according to system purposes. If the data is the core asset in a system, STPA-DFSec is suitable for analysis insecure issues with more considerations on the information. If providing proper and secure services is the main object of a system, STPA-Sec is applicable to identify insecure issues linking with application scenarios. The second principle is to consider who uses it. STPA-DFSec is suitable for designers who are responsible for technical structure and design, while STPA-Sec is more useful for ones who design the system functionalities and make more high-level decisions.

Actually, system security engineering is not able to ensure absolute security but provides a sufficient base of evidence that supports claims that the expected level of trustworthiness has been achieved \cite{ross2016nist}. The analysis in security engineering is also not able to be proven complete, and the analysis results normally depend on the analyst's knowledge and design emphasis. However, a proper systematic approach can help the analyst to be more confident in the analysis completeness \cite{young2014inside}. Proper guide words help to reduce the dependency on personal experience and subjective thinking and lead to objective and valid results with less effort. Although the case study in this paper represents the authors' understanding of the system, the analysis results are comparable and meaningful because both analyses were performed by the same group of analysts.

\section{Conclusion}

In this paper, we have proposed a data-flow-based approach for security analysis of information-critical systems based on the STPA framework to overcome STPA-Sec's limitations. The analyses of a vehicle digital key system by using both the STPA-DFSec and STPA-Sec have been presented and compared to show how to use the approaches and how well both approaches work on the same use case. 

We have found that the proposed STPA-DFSec focuses on data flows and can reveal more details in information security aspects, which are hard to be addressed in the STPA-Sec analysis, while the STPA-Sec analyzes systems from the perspective of applications and more concerns safety-related security issues. Besides, since STPA-based approaches were created for high-level decisions rather than tactical details \cite{young2014inside}, the proposed STPA-DFSec extends considerations into lower levels with technical details. Furthermore, as an extension of the STPA series, the proposed approach, together with other STPA-based approaches, can be used to co-design complex systems in multi-disciplines from high to low system levels under the unified STPA framework. Social aspects and human factors can be included in the analysis, which are excluded in traditional analysis approaches.

In the future, we will study more industry cases and conduct experiments with different groups of analysts to validate and refine the proposed approach in practices since we performed both analyses in this paper, which might influence the validity of analysis outcomes. Furthermore, we will formalize the analysis process and design tools to achieve analysis results automatically for higher working efficiency.


\ifCLASSOPTIONcaptionsoff
  \newpage
\fi


\bibliographystyle{IEEEtran}
\bibliography{ref}










\end{document}